\documentclass{article}
\usepackage{CJK}
\usepackage{ijcai18}
\usepackage{indentfirst}
\usepackage{amsmath}
\usepackage{url}
\usepackage{times}
\usepackage{comment}
\usepackage{xcolor}
\usepackage{soul}
\usepackage[utf8]{inputenc}
\usepackage[small]{caption}
\usepackage{amsmath}
\usepackage{txfonts}
\usepackage{amssymb}
\usepackage{graphicx}
\usepackage{booktabs}
\usepackage{comment}
\usepackage{subfigure}
\usepackage{extarrows}
\usepackage{multirow}
\usepackage{multicol}
\usepackage{slashbox}
\usepackage{times}
\usepackage[lined,ruled,vlined]{algorithm2e}
\newtheorem{definition}{Definition}
\newtheorem{theorem}{Theorem}

\title{Feature Propagation on Graph: A New Perspective to Graph Representation Learning}
\author{Biao Xiang, Ziqi Liu, Jun Zhou, Xiaolong Li\\
Ant Financial\\
\{xiangbiao.xb, ziqiliu, jun.zhoujun, xl.li\}@antfin.com}

\begin{document}

\maketitle

\begin{abstract}

  We study feature propagation on graph, an inference process involved in graph representation learning tasks.
  It's to spread the features over the whole graph to the $t$-th orders, thus to expand the end's features.
  The process has been successfully adopted in graph embedding or graph neural networks,
  however few works studied the convergence of feature propagation. Without convergence guarantees, it may lead to unexpected
  numerical overflows and task failures. In this paper, we first define the concept of feature propagation on graph formally,
  and then study its convergence conditions to equilibrium states. We further link feature propagation
  to several established approaches such as node2vec and structure2vec.
  In the end of this paper, we extend existing approaches from represent nodes to edges (edge2vec) and demonstrate
  its applications on fraud transaction detection in real world scenario. Experiments show that it is quite competitive.
\end{abstract}

\section{Introduction}
In this paper, we study the feature propagation on graph, which forms the building blocks in many graph representation learning tasks.
Typically, the graph representation learning tasks aim to learn a function $f(\mathcal{X}, \mathcal{G})$ to somehow utilize the additional
graph structure in space $\mathcal{G}$, compared with traditional learning tasks $f(\mathcal{X})$ by only considering each sample
independently. The successes of graph representation approaches~\cite{grover2016node2vec,dai2016discriminative,kipf2016semi,hamilton2017inductive}
have proven to be successful on citation networks~\cite{sen2008collective}, biological networks~\cite{zitnik2017predicting}, and
transaction networks~\cite{Liu:2017:PNN:3133956.3138827} that can be formulated in graph structures.

One major process of graph representation learning tasks involves the feature propagation over the graph up to $t$-th orders.
Those approaches define various propagation manners based on such as, adjacency matrices~\cite{belkin2002laplacian}, $t$-order
adjacency matrices~\cite{cao2015grarep}, expected co-occurency matrices~\cite{perozzi2014deepwalk}\cite{grover2016node2vec} by conducting random walks.
Recently, graph convolutional networks have shown their promising results on various datasets.
They rely on either graph Laplacians~\cite{kipf2016semi} or on carefully-designed operators like mean, max operators over
adjacency matrix~\cite{hamilton2017inductive}.

However, few of graph representation learning tasks study the propagation process used in their inference procedures.
For instance, GCN~\cite{kipf2016semi} or structure2vec~\cite{dai2016discriminative} implicitly involve this procedure
in the form $$H^{(t+1)} = \phi(A) H^{(t)}W,$$ where $H \in \mathbb{R}^{N,K}$ denotes the learned embeddings of $N$ nodes
in vector space $\mathbb{R}^K$, the $t$ denotes the $t$-th iteration, $\phi(.)$ defines the operator on adjacency matrix $A \in \{0,1\}^{N,N}$
given graph $\mathcal{G}=\{\mathcal{V}, \mathcal{E}\}$. This propagation process is parameterized by $W \in \mathbb{R}^{K,K}$.
This iterative propagation process essentially propagate and spread each node $i$'s signals to $i$'s $T$-th step neighborhood over the graph.
Without the careful designs of the process under certain conditions, the propagation could be under risk of numeric issues.

In this paper, we are interested in the convergence condition of the propagation process to equilibrium state~\cite{langville2006updating},
hopefully can help the understanding of existing literatures in this domain:
(1) we first formulate the generic
framework of feature propagation on graphs;
(2) we connect existing classic approaches such as node2vec~\cite{grover2016node2vec}, a random walk based
graph embedding approach, and structure2vec~\cite{dai2016discriminative}, a graph convolution based approach, to our feature propagation framework;
(3) we study the convergence condition of feature propagation over graph to equilibrium state with $T\to \infty$ by using theory of M-matrix~\cite{plemmons1977m}, which is quite
simple and easy to implement by gradient projection;
(4) we further extend the existing node representation approaches to
edge representation, i.e. we propose ``edge2vec'' and show its applications on fraud transaction detection in a real world transaction
networks, which is essentially important in any financial systems. More importantly, ``edge2vec'' can deal with multiple links (transaction among two accounts
over a time period) among two nodes, which is essentially different from traditional settings like recommender systems (the user $i$ could have only one rating $r_{ij}$ on the item $j$,
i.e. only one link among two nodes).

This paper is organized as follows. In section~\ref{sec:preliminary}, we sets up the preliminary of this paper, and propose pairs of general definitions for
feature expansion and feature propagation in a unified learning framework. In section~\ref{sec:typical_way}, we discuss a typical feature propagation way, and
propose the sufficient conditions for its convergence. In section~\ref{sec:relationship}, we explore the connection between feature propagation and two types
of graph representation approaches. We finally extend the node embedding to edge embedding, and demonstrated its effectiveness by conducting experiments on
fraud transaction detection in section~\ref{sec:edge} and section~\ref{sec:experiment} respectively.

\section{Preliminary}\label{sec:preliminary}
Suppose the graph is $\mathcal{G=(V,E)}$, where $\mathcal{V}=\{1,2,...n\}$ is the node set, $\mathcal{E}=\{e_1,e_2,...e_m\}$ is the
edge set, and the adjacency matrix is $A=[a_{ij}]_{n*n}$ where $a_{ij}=1$ if $(i,j)\in\mathcal{E}$, and 0 otherwise.
$D=diag\{d_1,d_2,...d_n\}$ is the degree matrix of graph where $d_i=\sum_{j=1}^n{a_{ij}}$ is the degree of node
$i$(or $[d_1,d_2,...d_n]=A\mathbf{e}$). The feature set for node is $\mathcal{X}=\{x_1,x_2,...x_n\}$ where
$x_i=[x_{i1},x_{i2},...x_{id}]^T$ responds to node $v_i$, and denotes its responding feature matrix as
$X=[x_1, x_2,....x_n]^T$. If exists, we denote the feature set for edge as $\mathcal{X}^{(e)}=\{x_1^{(e)},x_2^{(e)},...x_m^{(e)}\}$
where $x_i^{(e)}=[x^{(e)}_{i1},x^{(e)}_{i2},...x^{(e)}_{id^{(e)}}]^T$ responds to the edge $e_i$ and denote its responding feature
matrix as $X^{(e)}=[x_1^{(e)},x_2^{(e)},...x_m^{(e)}]^T$. If the label locates in node, we denote the label vector as
$Y=[y_1, y_2, ...y_n]^T$, and if the label locates in edge, we denote the label vector as $Y^{(e)}=[y^{(e)}_1,y^{(e)}_2,...y^{(e)}_m]^T$.

For the traditional learning tasks (with or without graph topology), the typical way to build the fitting model is as follows
$$
Y = f(X;\theta) \ \ \ or \ \ \ Y^{(e)} = f(X^{(e)};\theta).
$$
However, this way only utilizes the features of node or edge itself. In a context-aware perspective, the features of neighbor or the neighbor's neighbor may also be useful. For example, in a social network, assume that one person didn't fill her age, it may be hard to get this feature once we only utilize the features of herself; but if we utilize her neighbors' features, we may estimate this feature by averaging her neighbors' ages or take their median. We denote the expanded feature as $\widetilde{X}$ and $\widetilde{X^{(e)}}$ according to the raw feature $X$ and $X^{(e)}$ respectively. And call the expanded process from $X$ and $X^{(e)}$ to $\widetilde{X}$ and $\widetilde{X^{(e)}}$ as feature expansion. We define this concept as follows

\begin{definition}\label{def:feaexp}
  (Feature Expansion). Suppose the raw feature of graph are $X$ and $X^{(e)}$, responding to node and edge respectively, if
  $$
  \widetilde{X} = \mathcal{P}(X, X^{(e)};\theta_p)\ \ \ or\ \ \ \widetilde{X^{(e)}} = \mathcal{P}(X^{(e)}, X;\theta_p)
  $$
  then we call $\widetilde{X}$ or $\widetilde{X^{(e)}}$ as expanded features, and call the function $\mathcal{P}$ as feature expansion function.
\end{definition}

With expanded features, the fitting model will be
\begin{equation}
  \begin{aligned}
    Y&=f(\widetilde{X};\theta)=f(\mathcal{P}(X, X^{(e)};\theta_p);\theta)\ \ \ or\\
    Y^{(e)}&=f(\widetilde{X^{(e)}};\theta)=f(\mathcal{P}(X^{(e)}, X;\theta_p);\theta)
  \end{aligned}
\end{equation}
which contains two sets of parameters $\theta_p$ and $\theta$, where $\theta_p$ is parameters for feature expansion and $\theta$ is for
fitting the final label. And the learning framework with feature expansion is as follows
\begin{itemize}
  \item[1.]\small{Initialize parameters $\theta_p$ and $\theta$;}
  \item[2.]\small{Expand the raw feature $X$ to $\widetilde{X}$ by expansion function $\mathcal{P}(X, X^{(e)};\theta_p)$;}
  \item[3.]\small{Compute the prediction $\hat{Y}=f(\widetilde{X};\theta)$;}
  \item[4.]\small{Back propagate the $loss(Y,\hat{Y})$ to update $\theta$ and $\theta_p$;}
  \item[5.]\small{Repeat step 2-4 until $loss(Y,\hat{Y})$ minimized;}
\end{itemize}

In graph, the feature expansion is usually propagated via the graph topology, and the feature of node or edge is expanded by its neighbors in the $t$-th orders. Since this feature expansion process relies on the feature propagation through the graph topology, we call this process as feature propagation with definition as follows:
\begin{definition}\label{def:feaprop}
  (Feature Propagation). Suppose the raw feature of graph are $X=[x_1,x_2,...x_n]^T$ and $X^{(e)}=[x_1^{(e)},x_2^{(e)},...x_m^{(e)}]^T$, responding to node and edge respectively, if for each $i\in[1,n]$ and $j\in[1,m]$
  \begin{equation}
    \begin{aligned}
      \widetilde{x_i}&=\mathcal{P}(x_i, \mathop{\{\widetilde{x_k}\}}\limits_{\tiny{k\ is\ i's\ neighbor}}, \mathop{\{\widetilde{x_k^{(e)}}\}}\limits_{\tiny{e_k\ is\ adjoint\ to\ node\ i}};\theta_p)\ \ \ \ \ \ \ \ or\\
      \widetilde{x_j^{(e)}}&=\mathcal{P}(x_j^{(e)}, \mathop{\{\widetilde{x_k}\}}\limits_{\tiny{node\ k\ is\ related\ to\ e_j}};\theta_p)
    \end{aligned}
  \end{equation}
  , then we call $\widetilde{X}$ or $\widetilde{X^{(e)}}$ as propagation-expanded feature, and call the function $\mathcal{P}$ as feature propagation function.
\end{definition}

Although in feature propagation each node/edge only takes advantage of its neighbors' information, it still could get the information farther away through the iteratively propagation of definition~\ref{def:feaprop}.

In this section, we propose the general definitions for feature expansion and feature propagation in graph and propose the learning
framework with feature expansion. In the next section, we will discuss a typical feature propagation way, which has strong connection
with the recent popular graph representation learning method.

\section{A Typical Way for Feature Propagation}\label{sec:typical_way}
The typical way to expand node's features by propagation is as follows, which is a generalization of pagerank equation~\cite{page1999pagerank,xiang2013pagerank},
\begin{equation}
  \label{eq:typical_way}
  \widetilde{x_i}=W_1^T x_i + W_2^T \sum_{j\in \mathcal{N}(i)}\widetilde{x_j}\ \ \ for\ i=1,2,...n
\end{equation}
where $\mathcal{N}(i)$ is the neighbor set of node $i$, $W_1=[w_{ij}^{(1)}]_{d*d'}$ and $W_2=[w_{ij}^{(2)}]_{d'*d'}$ are the parameters of Eq.~\ref{eq:typical_way}(thus, the dimension of $\widetilde{x_i}$ is $d'$). For the convenience, we call $W_2$ as propagation matrix in this paper.
This equation group could be rewritten as
\begin{equation}\label{eq:typical_way_2}
  \widetilde{X} = X W_1 + A \widetilde{X} W_2
\end{equation}

Breaking up the group of equations, we have
\begin{equation}\label{eq:typical_way_expand}
\begin{aligned}
  \widetilde{x_{ij}}&=\sum_{k=1}^d{x_{ik}w_{kj}^{(1)}}+\sum_{p=1}^n\sum_{q=1}^{d'}{a_{ip}\widetilde{x_{pq}}w_{qj}^{(2)}}\\
\ \ \ &for\ i=1,2,...n\ and\ j=1,2,...d'
\end{aligned}
\end{equation}
And let $s=i*n+j$, $Z=[z_s]_{1*(n*d')}\ with\ z_s=\widetilde{x_{ij}}$, $L=[l_s]_{1*(n*d')}\ with\ l_s=\sum_{k=1}^d{x_{ik}w_{kj}^{(1)}}$ and
  $A'=[a_{st}^{'}]_{(n*d')*(n*d')}$ with
\begin{equation}
  \label{eq:A_prime}
  \begin{aligned}
    a_{st}^{'}&=a_{ip}w_{qj}^{(2)}=\left\{
            \begin{aligned}
                &0\ &if\ a_{ip}=0\\
                &w_{qj}^{(2)}\ &otherwise
            \end{aligned}
            \right.,\ \ \ with\\ s&=i*n+j,\ t=p*n+q
  \end{aligned}
\end{equation}
Equation~\ref{eq:typical_way_expand} could be rewritten as
$$
z_s = l_s + \sum_{t=1}^{n*d'}{z_t}*a_{st}^{'}\ \ \ for\ s=1,2,...n*d'
$$
After summing up, it becomes
$$
Z = L + A'Z
$$
If matrix $(I-A')$ is invertible, we will get
\begin{equation}\label{eq:typical_way_solution}
  Z = (I-A')^{-1}L
\end{equation}
However, $(I-A')$ is not invertible naturally, we should set some conditions to make it be. From Equation~\ref{eq:A_prime}, we could see that, only propagation matrix $W_2$ will affect the invertibility of $(I-A')$.

From the theory of M-matrix, if $(I-A')$ satisfies the following two conditions, it will be invertible.
\begin{itemize}
  \item[1.] $A'\geq 0$, which, by Eq.~\ref{eq:A_prime}, is equivalent to $W_2$ should be a nonnegative matrix;
  \item[2.] $A'\mathbf{e}< \mathbf{e}$, which, with the derivation in footnote\footnote{let's denote $W_2^T\mathbf{e}=[w_1,w_2,...w_{d'}]^T$, from Eq.~\ref{eq:A_prime},
      \begin{equation}\nonumber
        \begin{aligned}
          A'\mathbf{e}< \mathbf{e} &\iff \forall s, \sum_{t=1}^{n*d'}a_{st}^{'}<1 \iff \forall i,j, \sum_{p=1}^{n}\sum_{q=1}^{d'}a_{ip}w_{qj}^{(2)} < 1 \\ &\iff  \forall i,j, \sum_{p=1}^{n}a_{ip}\sum_{q=1}^{d'}w_{qj}^{(2)} < 1 \iff  \forall i,j, d_i w_j < 1\\&\iff \max\{w_1,w_2,...w_{d'}\}< 1/{\max\{d_1,d_2,...d_n\}}
        \end{aligned}
      \end{equation}}, is equivalent to $\max\{W_2^T\mathbf{e}\}< 1/{\max\{d_1,d_2,...d_n\}}$
\end{itemize}
However, condition 2 is a very demanding condition. If there exists a node with very large degree, the row sum of $W_2$ will have to be very small. To solve this issue, we could make the below changes to Eq.~\ref{eq:typical_way_2}, i.e. replace the matrix $A$ as $D^{-1}A$. Then, Eq.~\ref{eq:typical_way_2} changes to
\begin{equation}\label{eq:typical_way_3}
  \widetilde{X} = X W_1 + D^{-1} A \widetilde{X} W_2
\end{equation}
Dive into each $\widetilde{x_i}$, we have
\begin{equation}\label{eq:typical_way_4}
  \widetilde{x_i}=W_1^T x_i + W_2^T \sum_{j\in \mathcal{N}(i)}\frac{1}{d_i}\widetilde{x_j}
\end{equation}

Under this feature propagation process, the above condition 2 will change to $\max\{W_2\mathbf{e}\}\le 1$. Comparatively, this condition is easier to be guaranteed.

Summing up the above derivations, we form the following theorem.
\begin{theorem}
  \label{th:propagation_matrix}
  For feature propagation method as Eq.~\ref{eq:typical_way_3} or \ref{eq:typical_way_4}, the propagation matrix $W_2$ if satisfy the following conditions, the propagation process will be convergent.
  \begin{itemize}
    \item{condition 1.} $W_2$ is nonnegative.
    \item{condition 2.} $\max\{W_2^T\mathbf{e}\}< 1$.
  \end{itemize}
\end{theorem}

Theorem~\ref{th:propagation_matrix} proposed a pair of sufficient conditions to guarantee the convergence of feature propagation, but they are not necessary conditions. When the propagation matrix $W_2$ satisfies the conditions in theorem~\ref{th:propagation_matrix}, the feature propagation process as Eq.~\ref{eq:typical_way_3} will be convergent. Otherwise, the feature expansion may lead to explode which actually has been confirmed by the practical experiences.

\section{Relationship to Graph Representation Learning}\label{sec:relationship}
Recent years have seen a surge of research on graph representation and node embedding. These works could be roughly categorized into two types: 1)
embeddings with graph structure only~\cite{perozzi2014deepwalk,grover2016node2vec,abu2017watch}, and 2) embeddings with both structure and features
(or attributes)~\cite{kipf2016semi,dai2016discriminative,hamilton2017inductive}. In this section, we discuss the relationship
between feature propagation and graph representation.

\subsection{With Graph Structure Only}\label{sec:relationship1}
For the typical feature propagation way as Eq.~\ref{eq:typical_way_3}, if we let each node feature $x_i$ as a one-hot vector\footnote{which means node $i$ contains no feature, but only its identity.} (i.e. $X=I$), $W_1$ as a randomly initialized matrix $C=[c_1,c_2,...c_{d'}]=[c_{ij}]_{n*d'}$, $W_2=\alpha I\ (\alpha< 1)$ (must satisfy the two conditions in Theorem~\ref{th:propagation_matrix}), and denote $T=D^{-1}A~$\footnote{traditionally, $T$ is called as transition matrix, there is $T\mathbf{e}=\mathbf{e}$}, then 
Eq.~\ref{eq:typical_way_3} will be
\begin{equation}\label{eq:emb_structure0}
  \widetilde{X}= C + \alpha T\widetilde{X}.
\end{equation}
If substituting the above equation into its left side recursively, we will get
\begin{equation}\label{eq:emb_structure}
  \widetilde{X}=(I+\alpha T + \alpha^2 T^2 + ...)C.
\end{equation}
Let's denote
$$P = [p_1,p_2,...p_n]^T = [p_{ij}]_{n*n} = \lim_{k\rightarrow \infty}\sum_k{\alpha^k T^k}.$$
Because $\alpha<1$, the infinite sequence of $P$ will be converged gradually. Approximately, $T^k$ is the $k$-step transition probability matrix between any pair of nodes. Thus, $P$ is the weighted sum of $k$-step transition matrix with weight $\alpha^k$ and we call $P$ as proximity matrix. Its entry $p_{ij}$ depicts the transition probability from node $i$ to node $j$ by 0-step, 1-step, up to $\infty$-steps, and $p_i$ depicts the transition probability from node $i$ to any node in the graph. Thus, if node $i$ and node $j$ close to each other in the graph, $p_i$ and $p_j$ will be close too. From Eq.~\ref{eq:emb_structure}, we have
$$
\widetilde{x}_i = [p_ic_1,p_ic_2,...p_ic_{d'}],
$$
then
$$
\widetilde{x}_i-\widetilde{x}_j = [(p_i-p_j)c_1,(p_i-p_j)c_2,...(p_i-p_j)c_{d'}].
$$
If node $i$ is close to node $j$ in graph (which means $p_i$ is close to $p_j$\footnote{for $p_{ii}$ must be larger than $p_{ji}$ and $p_{jj}$ must be larger than $p_{ij}$, it will impact the comparison between $p_i$ and $p_j$. The better way is to adjust $P$ as $\hat{P}=P-I$ or adjust $\widetilde{X}$ as $\hat{X}=\hat{P}C=(\alpha T + \alpha^2 T^2 + ...)C=\widetilde{X}-C$.}), then $\widetilde{x}_i-\widetilde{x}_j$ will close to 0 no matter how
the $C$ is initialized. In~\cite{abu2017watch}, the authors revisited DeepWalk~\cite{perozzi2014deepwalk} and GloVe~\cite{pennington2014glove}, and find that their proximity matrices are:
\begin{equation}\label{eq:deepwalk}
  \begin{aligned}
    P^{\mathrm{DeepWalk}[K]}&=\sum_{k=1}^K[1-\frac{k-1}{K}]T^k,\\
    P^{\mathrm{GloVe}[K]}&=\sum_{k=1}^K[\frac{1}{k}]T^k,
  \end{aligned}
\end{equation}
respectively. Compared with the two proximity matrices above, the major differences between ours $P$ and theirs is the
decay weight of $T^k$. And our weight $\alpha^k$ is as reasonable as $1-(k-1)/K$ or $1/k$. Thus, $\widetilde{X}$ is a
reasonable first type embedding.

\begin{figure*} [th]
  \begin{center}
    \subfigure{\includegraphics[width=44mm,height=30mm]{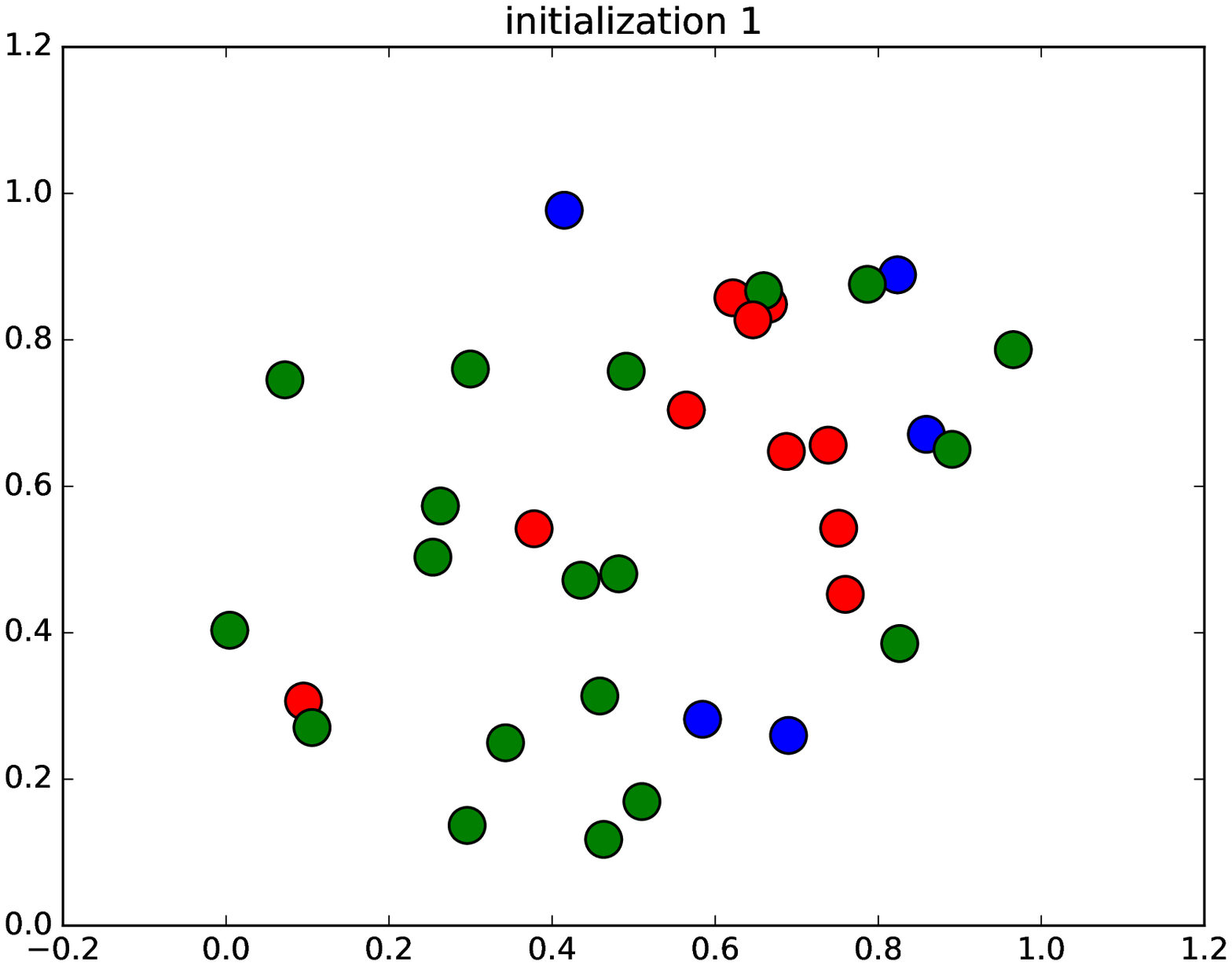}}\hspace{-3.5mm}
    \subfigure{\includegraphics[width=44mm,height=30mm]{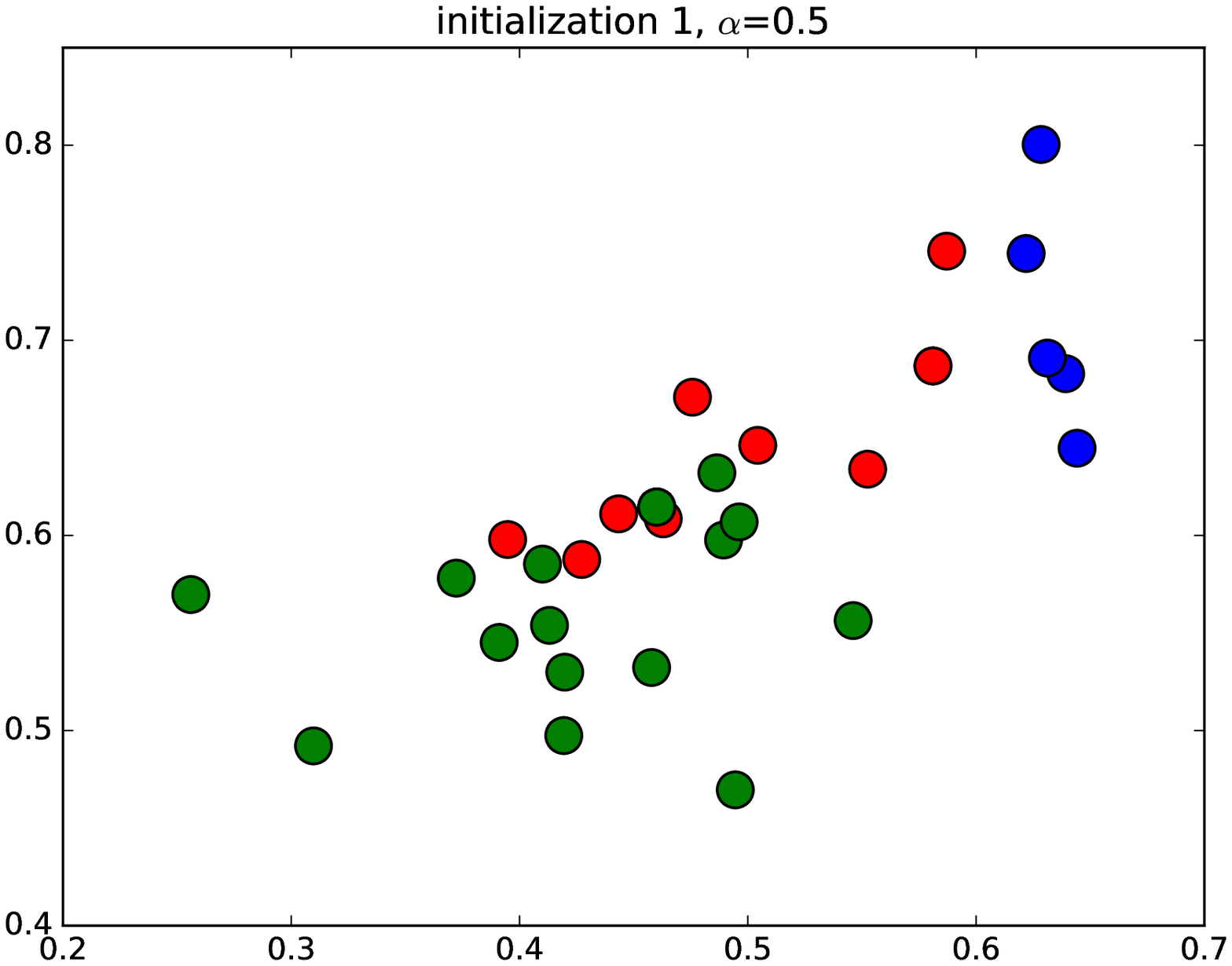}} \hspace{-3.5mm}
    \subfigure{\includegraphics[width=44mm,height=30mm]{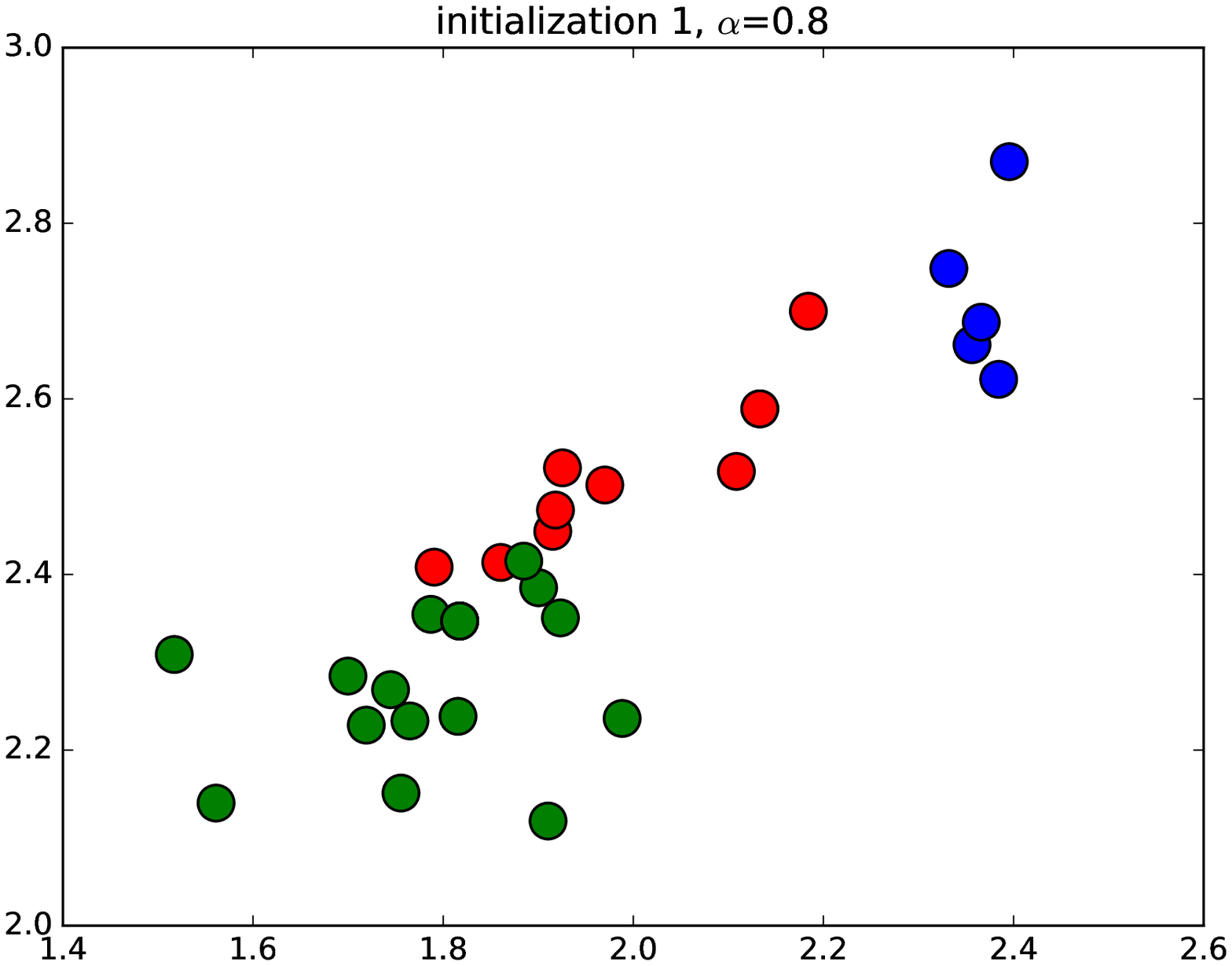}}\hspace{-3.5mm}
    \subfigure{\includegraphics[width=44mm,height=30mm]{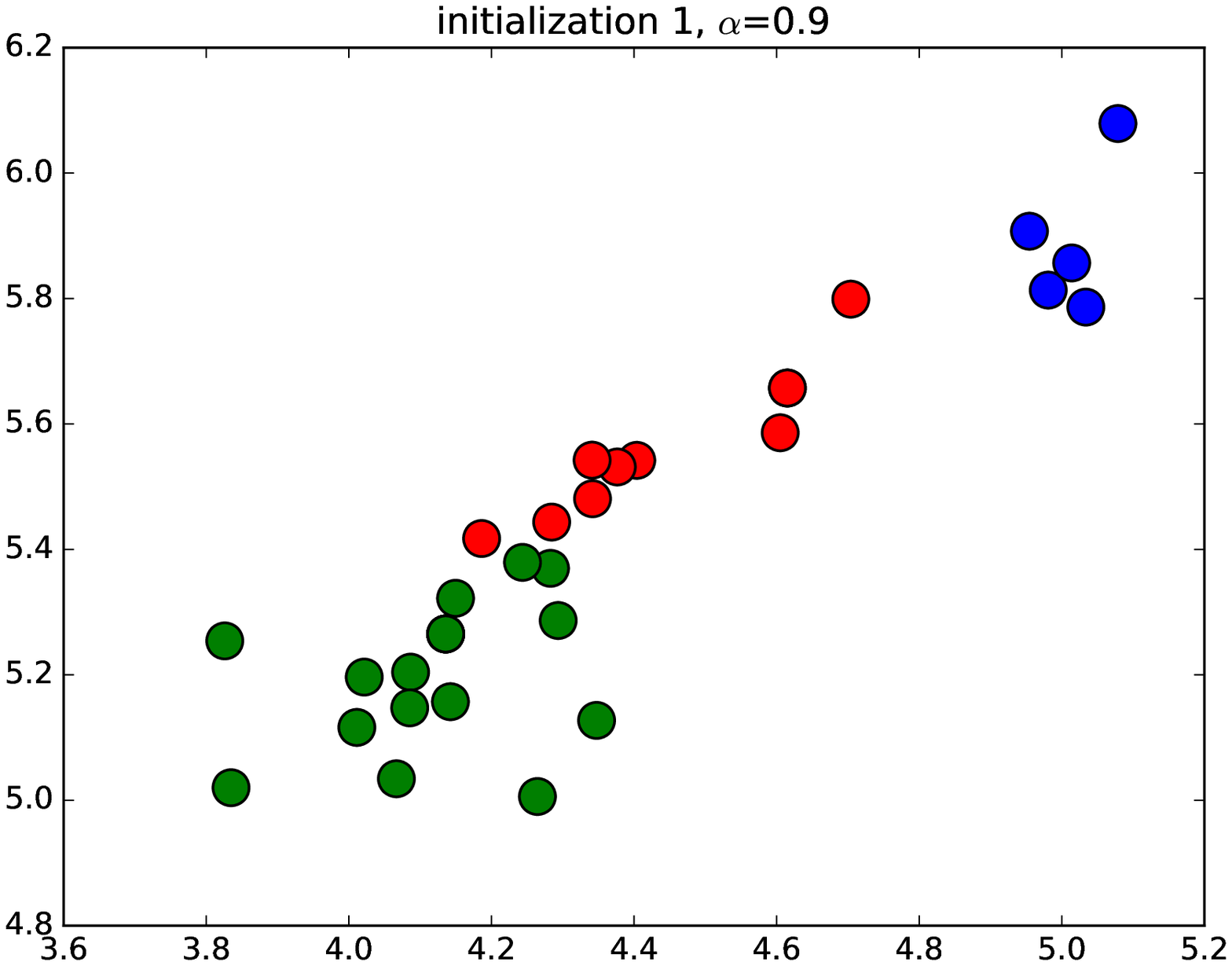}}\\ \vspace{-3.5mm}
    \subfigure{\includegraphics[width=44mm,height=30mm]{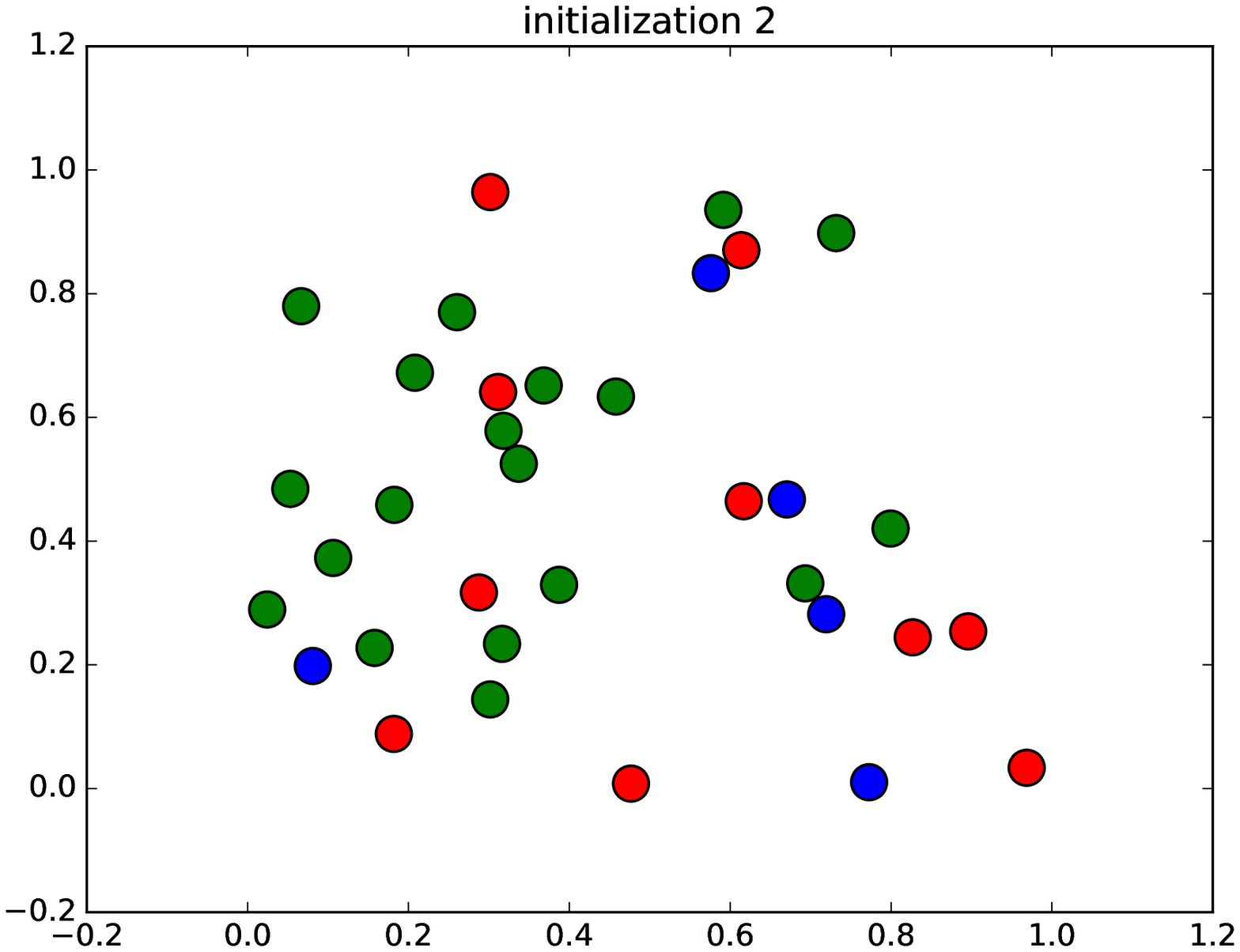}}\hspace{-3.5mm}
    \subfigure{\includegraphics[width=44mm,height=30mm]{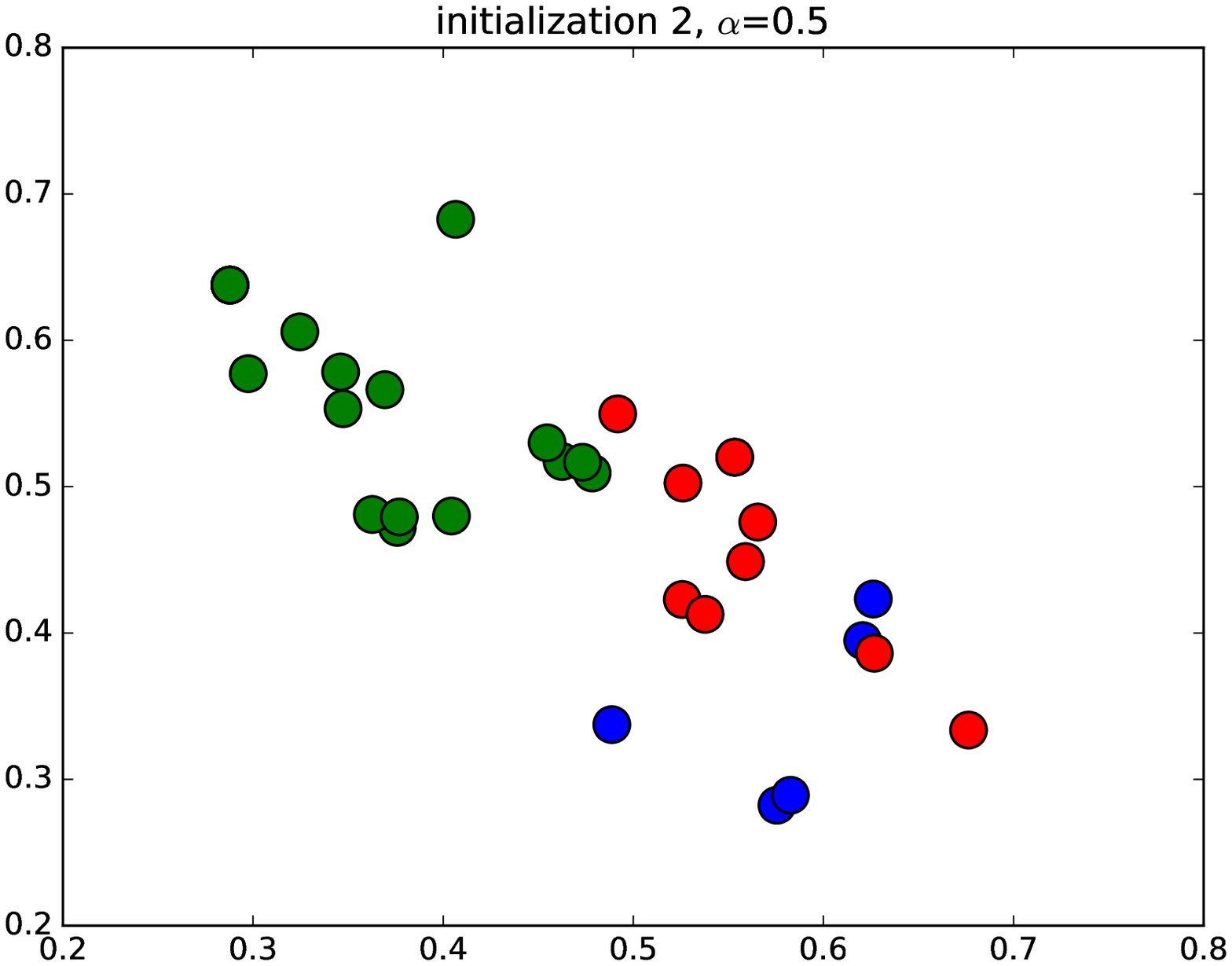}}\hspace{-3.5mm}
    \subfigure{\includegraphics[width=44mm,height=30mm]{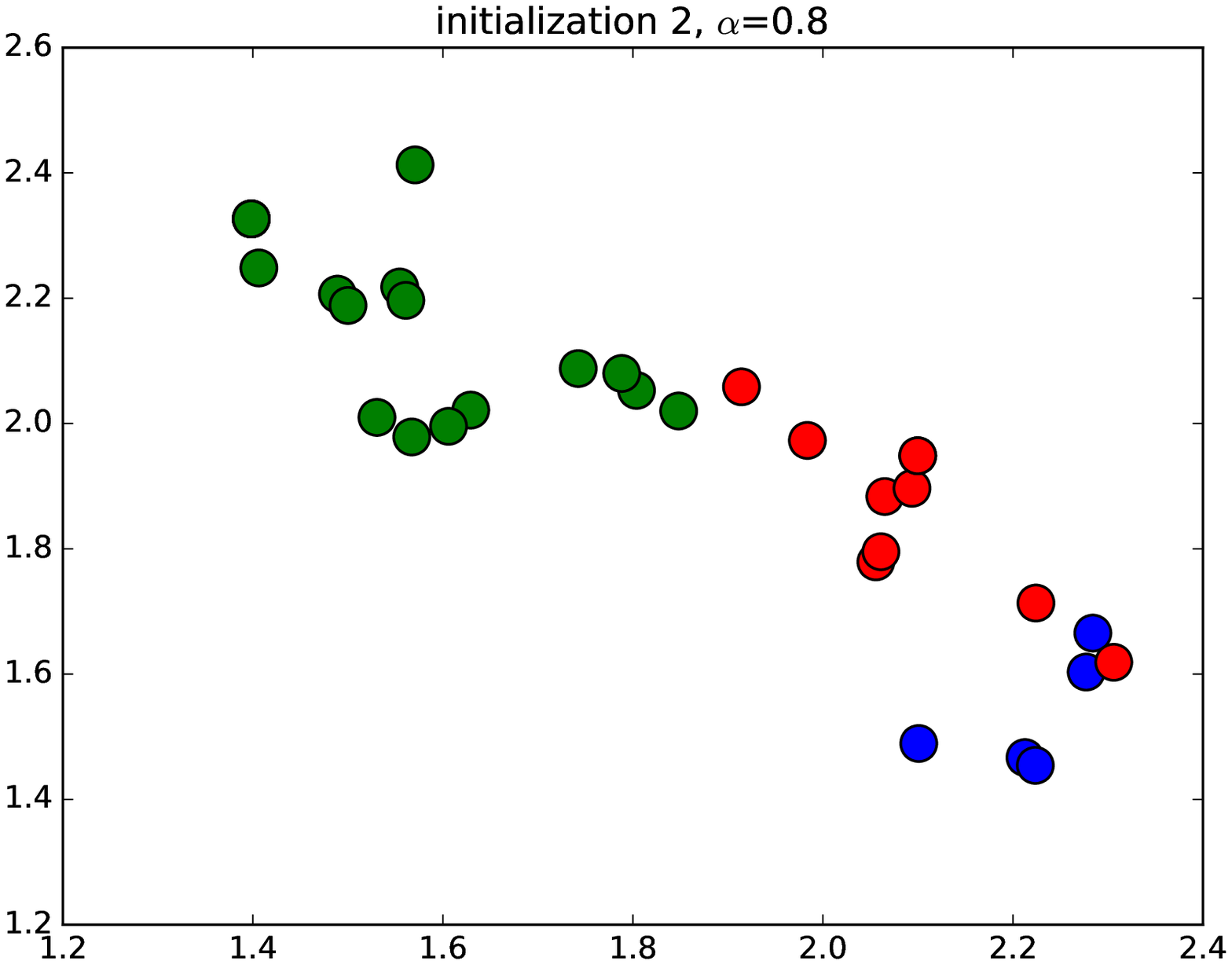}}\hspace{-3.5mm}
    \subfigure{\includegraphics[width=44mm,height=30mm]{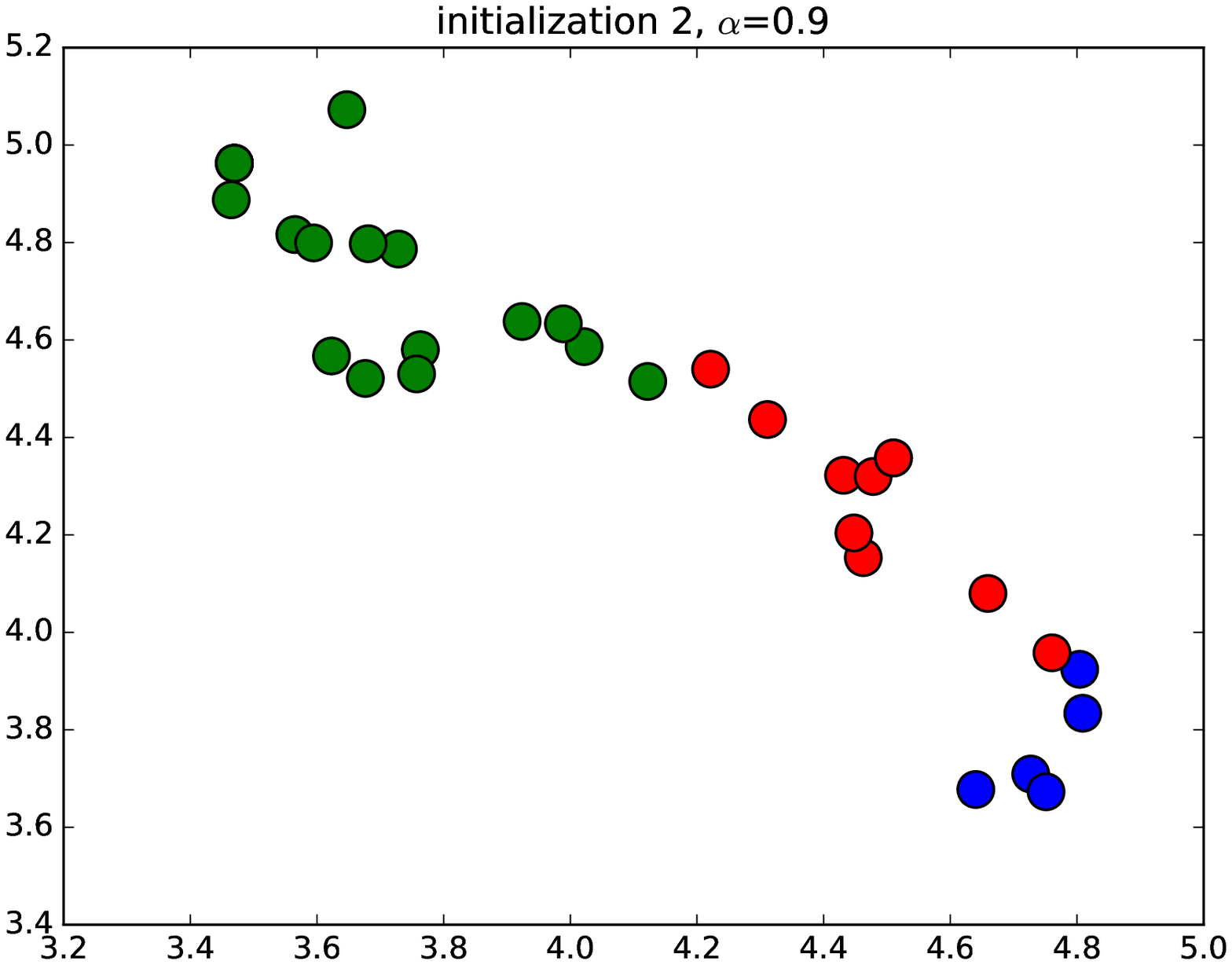}}\\ \vspace{-3.5mm}
    \subfigure{\includegraphics[width=44mm,height=30mm]{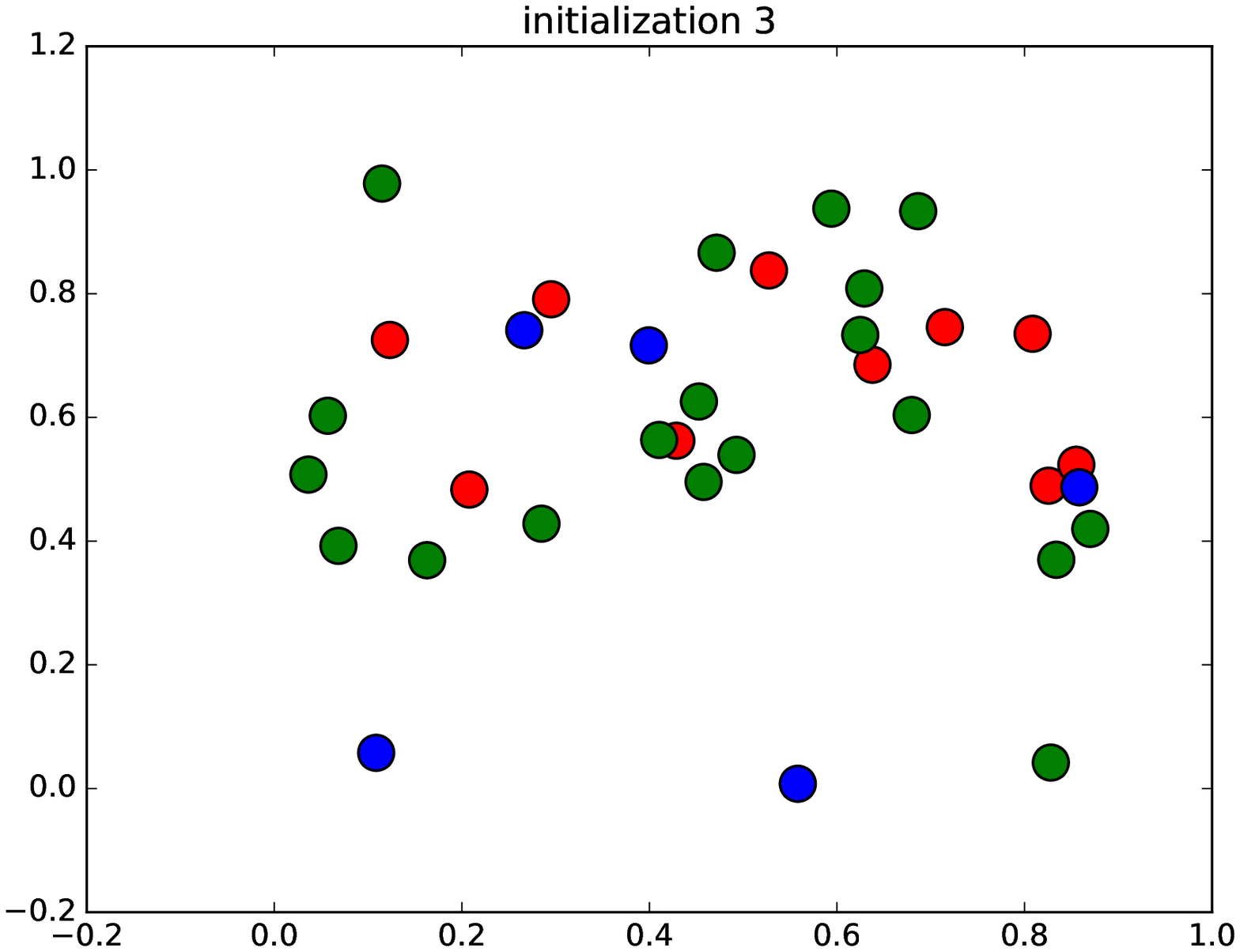}}\hspace{-3.5mm}
    \subfigure{\includegraphics[width=44mm,height=30mm]{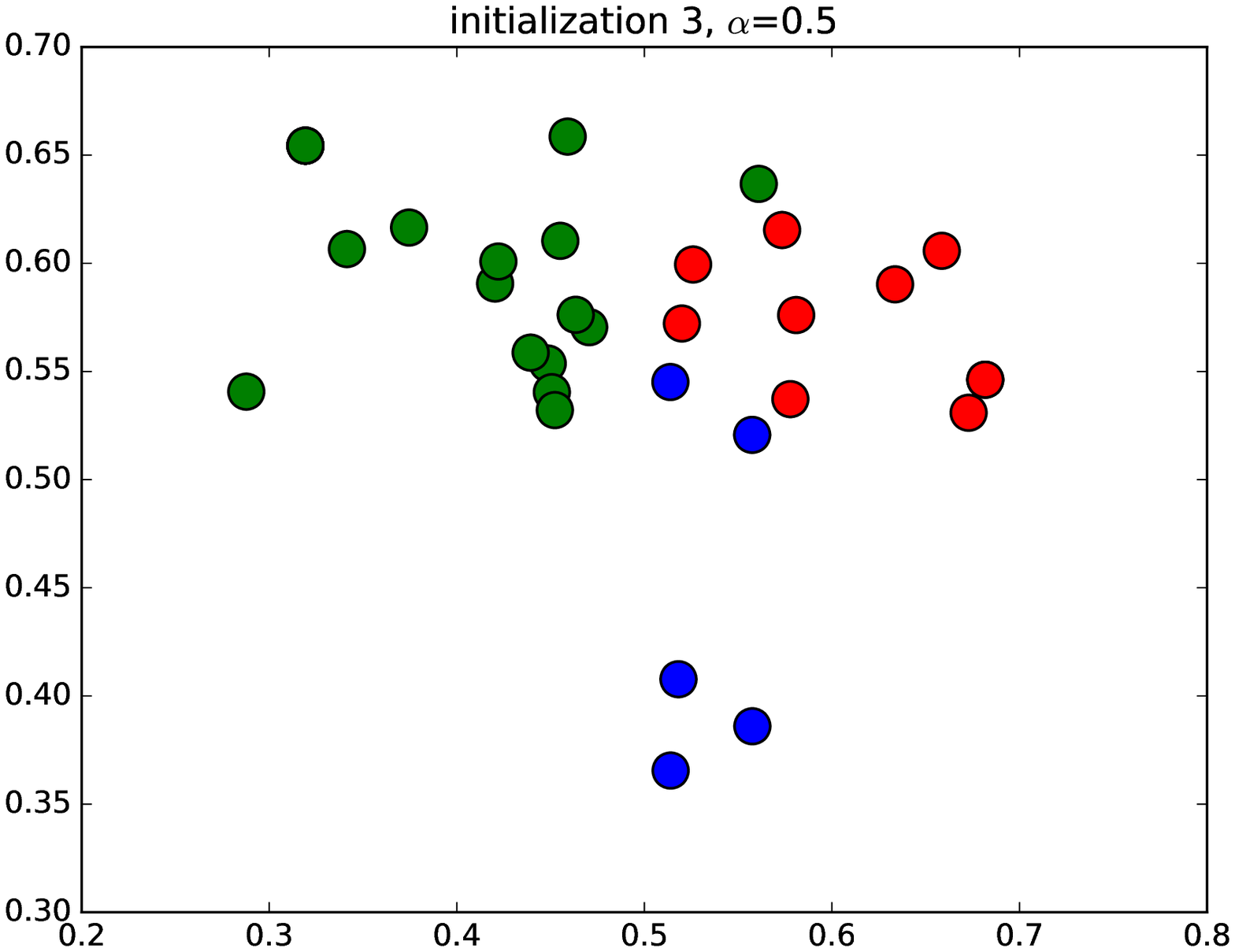}}\hspace{-3.5mm}
    \subfigure{\includegraphics[width=44mm,height=30mm]{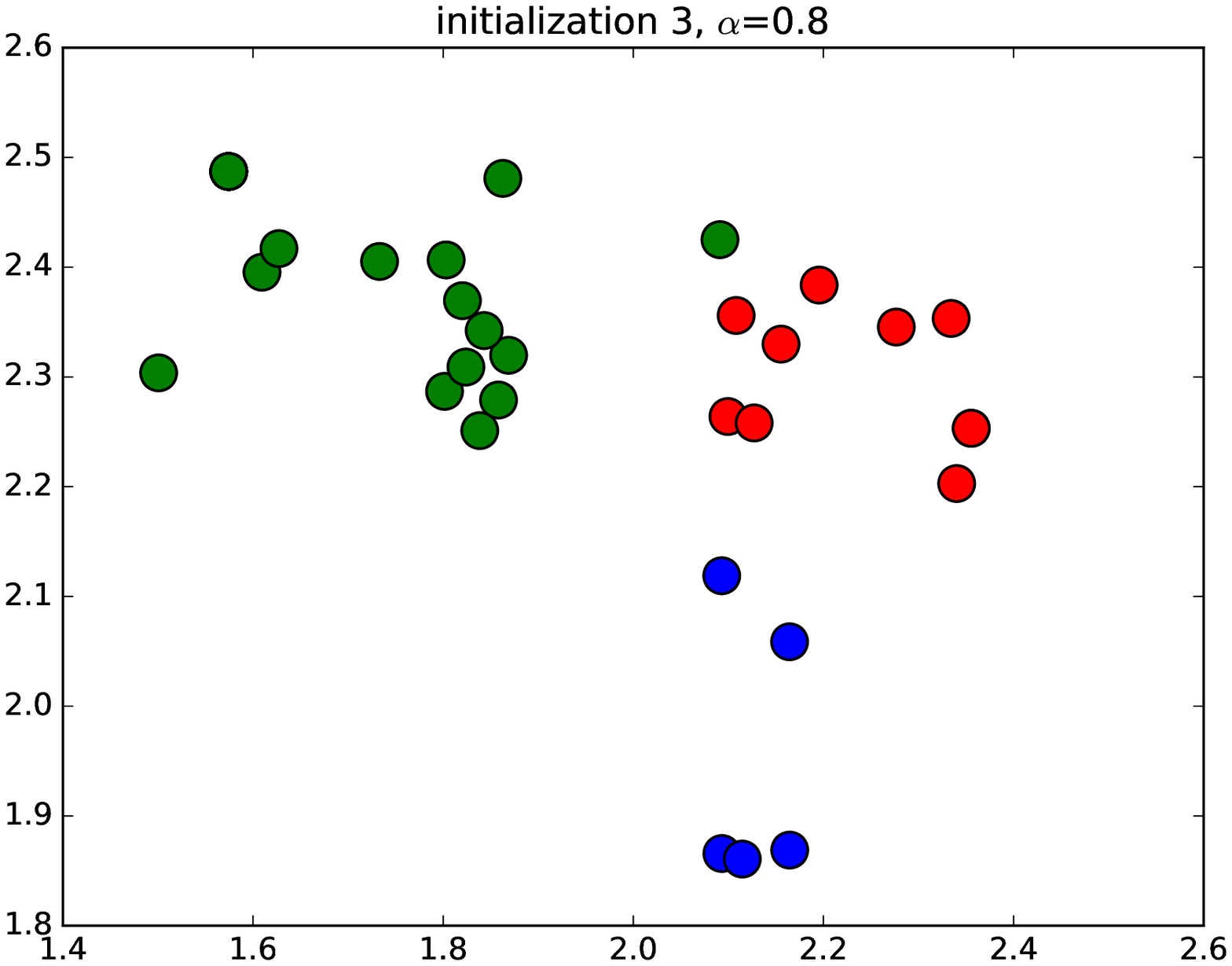}}\hspace{-3.5mm}
    \subfigure{\includegraphics[width=44mm,height=30mm]{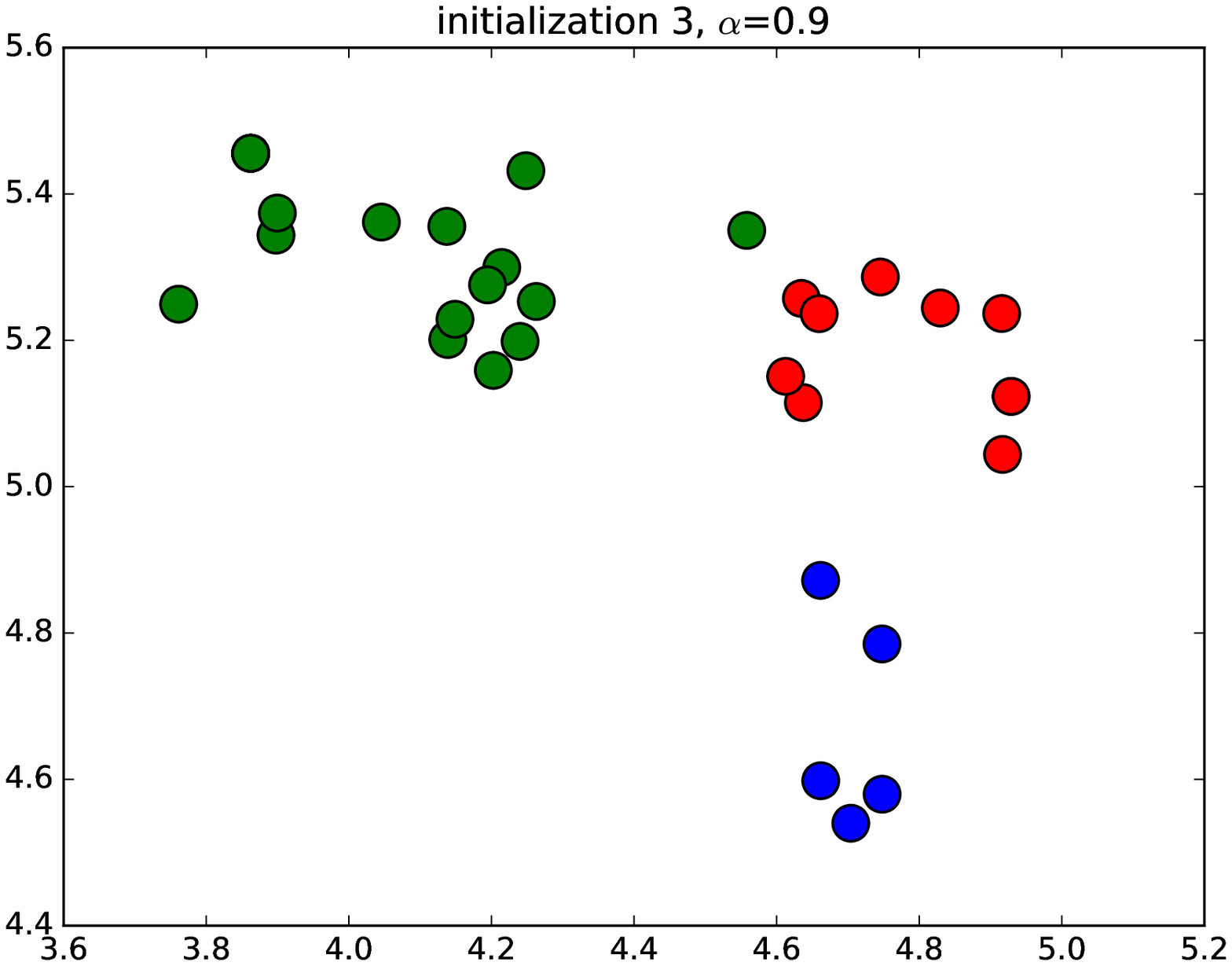}}\\
  \end{center}
  \vspace{-0.6cm}
  \caption{Node embedding for Zachary Karate Club social network. Different color indicates different community. Each row of figures is a group of results with the same initialization of $W_1$ and different $\alpha\in\{0.5,0.8,0.9\}$.}
  \label{fig:zachary}
 \vspace{-3.5mm}
\end{figure*}
Figure~\ref{fig:zachary} displays an example embedding for the famous Zachary Karate Club social network~\cite{perozzi2014deepwalk}, where we use two dimensional node embeddings
to capture the community structure implicit in the social network. We changed the initialization of $W_1$ and $\alpha$ in $W_2$, and could see that:
\begin{itemize}
  \item[1.] no matter how the $W_1$ is initialized, the embeddings can capture the community structures in the network pretty well;
  \item[2.] as the propagate parameter $\alpha$ becomes larger, the nodes in a community will tend to aggregate;
\end{itemize}
Feature propagation as Eq.~\ref{eq:typical_way_3} could be a simple way of the first type embedding when $W_1$ and $W_2$ satisfy center conditions.

\subsection{With Structure and Features on Graph Simultaneously}\label{sec:relationship2}
In structure2vec~\cite{dai2016discriminative}, a graph convolution based approach, the node embedding was formulated as
$$
\widetilde{\mu}_i=\sigma(W_1 x_i + W_2\sum_{j\in\mathcal{N}(i)}\widetilde{\mu}_j+W_3\sum_{j\in\mathcal{N}(i)}x_j)
$$
where $\sigma:=max\{0,\cdot\}$ is a rectified linear unit function. Suppose the dimension of $\widetilde{\mu}_i$ is $d'$, i.e. $\widetilde{\mu}_i=[\widetilde{\mu}_{i1},\widetilde{\mu}_{i2},...\widetilde{\mu}_{id'}]$. Using the similar derivations in section~\ref{sec:typical_way}, we can get

\begin{equation}~\label{eq:emb_s1}
\begin{aligned}
  \widetilde{\mu}_{ij}&=\sigma(c_{ij}+\sum_{p=1}^n\sum_{q=1}^{d'}{a_{ip}\widetilde{\mu}_{pq}w_{qj}^{(2)}})
  \\ &for\ i=1,2,...n,\ and\ j=1,2,...d'
\end{aligned}
\end{equation}
where $c_{ij}=\sum_{k=1}^d w^{(1)}_{jk}x_{ik}+\sum_{p=1}^n\sum_{q=1}^{d'}{a_{ip}x_{pq}w^{(3)}_{qj}}$. Without loss of generality,
let's suppose there are $K$ variables $[\widetilde{\mu}_{{i_1}{j_1}},\widetilde{\mu}_{{i_2}{j_2}},...\widetilde{\mu}_{{i_K}{j_K}}]\equiv \mu$ in Eq.~\eqref{eq:emb_s1} are nonzeros,
while the other $n*d'-K$ variables are equal to $0$s. Then, Eq.~\eqref{eq:emb_s1} could be rewritten as
\begin{equation}\label{eq:emb_s2}
\begin{aligned}
  \widetilde{\mu}_{{i_s}{j_s}} = c_{ij}+\sum_{p=1}^n\sum_{q=1}^{d'}{a_{i_sp}\widetilde{\mu}_{pq}w_{qj_s}^{(2)}}\\\ \ \ \ \  for\ s=1,2,...K
\end{aligned}
\end{equation}
This equation also could be resolved by the similar derivations in section~\ref{sec:typical_way}. The final solution is in the form of
$$
\mu = (I-A'')^{-1}C
$$
where $C=[c_{ij}]$ and $A''$ is the matrix $A'$ in section~\ref{sec:typical_way} after removing $n*d'-K$ corresponding rows and columns.

Similarly, if we want the node embeddings converge and do not get explode, the matrix $W_2$ also needs to satisfy certain conditions like in section~\ref{sec:typical_way}. The relu function $\sigma$ decreased the scale of equations, but it hasn't changed the essence of linear system.

\section{Extension to Edge}\label{sec:edge}
The above section discussed the feature propagation when the graph only contains node features(i.e. $X$). However, in the many real scenarios, the graph may contains edge features(i.e. $X^{(e)}$) too. If we neglect the edge features, it may weaken the model's performance. What's more, the label may locate in edge directly, we have to utilize the edge features especially when
there exits multiple links between two nodes. This section we will discuss the feature propagation when the graph contains edge features in multiple-links settings.

For each edge $e_i$, suppose $i_s$ and $i_t$ are source node and target node of $e_i$ respectively, in mathematical form,  i.e. $e_i=(i_s,i_t)$. Suppose $\mathcal{S}(k)$ is the set of edge which takes node $k$ as source node and $\mathcal{T}(k)$ is the set of edge which takes node $k$ as target node. Suppose
\begin{equation}
  \begin{aligned}
    C_s&=[c^{(s)}_{ij}]_{m*n},\ where\  c^{(s)}_{ij}=\left\{\begin{aligned}
      &1\ \ &if\ e_i\in \mathcal{S}(j)\\
      &0\ \ &otherwise
    \end{aligned}\right.\\
    C_t&=[c^{(t)}_{ij}]_{m*n},\ where\  c^{(t)}_{ij}=\left\{\begin{aligned}
      &1\ \ &if\ e_i\in \mathcal{T}(j)\\
      &0\ \ &otherwise
    \end{aligned}\right.\\
  \end{aligned}
\end{equation}
and we call $C_s$ and $C_t$ as source incidence matrix and target incidence matrix respectively. Obviously, there is
\begin{equation}
\label{eq:incidence_equation}
\begin{aligned}
  C_s^{T}\centerdot C_s=I,\ \ \ C_s^{T}\centerdot C_t=0\\
  C_t^{T}\centerdot C_s=0,\ \ \ C_t^{T}\centerdot C_t=I
\end{aligned}
\end{equation}

We could expand the features by the following way
\begin{equation}
\left\{
\begin{aligned}
  &\widetilde{x_i^{(e)}} &=& W_1^T x_i^{(e)} + W_2^T \widetilde{x_{i_s}} + W_3^T \widetilde{x_{i_t}}\\
  &\widetilde{x_j} &=& W_4^T x_j + W_5 \sum_{k\in \mathcal{N}(j)}\frac{1}{d_j}\widetilde{x_k} + W_6^T \sum_{e_s\in \mathcal{S}(j)} \widetilde{x_s^{(e)}} + W_7^T \sum_{e_t\in \mathcal{T}(j)} \widetilde{x_t^{(e)}}
\end{aligned}\right.
\end{equation}
which could be rewritten as
\begin{equation}
\label{eq:feaprop_all}
\left\{
\begin{aligned}
  &\widetilde{X^{(e)}}&=&X^{(e)}W_1 + C_s\widetilde{X}W_2 + C_t\widetilde{X}W_3\\
  &\widetilde{X}&=&XW_4+D^{-1}A\widetilde{X}W_5+AC_s^T\widetilde{X^{(e)}}W_6+AC_t\widetilde{X^{(e)}}W_7
\end{aligned}\right.
\end{equation}
combine Eq.~\ref{eq:incidence_equation} and Eq.~\ref{eq:feaprop_all}, we will get
\begin{equation}
\label{eq:feaprop_solution}
  \begin{aligned}
    \widetilde{X}=&(XW_4+AC_s^TX^{(e)}W_1W_6+AC_t^TX^{(e)}W_1W_7)+\\
    &D^{-1}A\widetilde{X}(W_5+W_2W_6+W_2W_7)
  \end{aligned}
\end{equation}
Follow the similar derivations in section~\ref{sec:typical_way}, we could know that if we want the feature propagation to be convergent, we should satisfy certain conditions. Based on theorem~\ref{th:propagation_matrix}, the following two conditions can guarantee the convergence of the above feature propagation process:
\begin{itemize}
  \item[1.] $(W_5+W_2W_6+W_2W_7)$ should be nonnegative;
  \item[2.] $\max\{(W_5+W_2W_6+W_2W_7)^T\mathbf{e}\}< 1$
\end{itemize}
The above condition 2 is not easy to be satisfied since it depends on the interaction between different matrices. And, the feature propagation with edge features on graph is easy to explode. To eliminate this obstacle, we could simplify the feature propagation process by setting $W_6$ and $W_7$ to be 0, which means the expanded features $\widetilde{X}$ on nodes won't rely on features of edges. Then, we have the following feature propagation equation:
\begin{equation}
\label{eq:feaprop_reduce}
\left\{
\begin{aligned}
  &\widetilde{X^{(e)}}&=&X^{(e)}W_1 + C_s\widetilde{X}W_2 + C_t\widetilde{X}W_3\\
  &\widetilde{X}&=&XW_4+D^{-1}A\widetilde{X}W_5
\end{aligned}\right.
\end{equation}
We call this feature propagation way for edge as edge2vec in this paper. For edge2vec, we only need to guarantee the following conditions
\begin{itemize}
  \item[1.] $W_5 > 0$;
  \item[2.] $\max\{W_5^T\mathbf{e}\} < 1$;
\end{itemize}
to guarantee its convergence.

\section{Applications in Fraud Transaction Detection}\label{sec:experiment}
In the previous sections, we first proposed the concept ``feature propagation'' in a unified framework. We link feature propagation
as a basic building block to several graph representation tasks, and point out that the convergence conditions involved in generic graph
representation tasks. We further propose a simple extension of feature propagation to edge2vec where features and labels located on
edges. In this section, we conduct experiments on real world data to demonstrate the performance of edge2vec and its convergence.


\subsection{DataSet}

In this section, we study a real world data at a leading casheless payment platform in the world, served more than hundred
millions of users. As a financial services provider, one of major problems faced is the risk control of fraudulent transactions. Detecting and identifying
the risk of fraud for each transaction plays the fundamental importance of the platform.

In particular, we study the fraud transaction in the online shopping setting, where sellers sell fake items to customers to reap undeserved profits.
Independently considering each transaction between a seller and a buyer cannot characterize useful information from the whole transaction network.
Considering the problem in the feature propagation framework over graph can help us understand underlying aggregation pattern
of the fraudulent transactions.

The experimental fraud transaction data\footnote{the data is randomly sampled over a time period with complete data desensitization (no personal profile, no user id).}
contains three types of features: 1) buyer's features 2) seller's feautres and 3) characterizations on each transaction. We treat each buyer and seller as a node of the
graph, and each transaction is an edge between buyer and seller. If one transaction $e_i$ is fraud, we label its corresponding edge as $y^{(e)}_i=[1,0]$,
otherwise label the transaction $e_i$ as $y^{(e)}_i=[0,1]$. Our task is to predict whether or not one edge is a fraud. The detailed statistics of the
data is described in Table~\ref{tab:data}.

Note that there could be multiple edges between a seller and a buyer, thus make the setting a bit different from traditional recommendation setting.
Our edge2vec can embed each edge into a vector space, so that it can help us to infer the risk of each edge in the graph.


\begin{table}[!th]
\centering
\caption{Fraud Transaction Detection Data Description.} \label{tab:data}
\begin{small}
\begin{tabular}{ccccc}
    \toprule
 &\#Nodes&\#Edges&\#Fraud&\#Normal \\
 \midrule
Training Data &626,003& 1,720,180& 31,737 &1,688,441\\
Testing Data & 1,355,824&4,034,962 &86,721 &3,948,241 \\
\bottomrule
\end{tabular}
\end{small}
\end{table}

\subsection{Treatment and Control Groups}
As discussed in section~\ref{sec:preliminary}, the learning framework is
$$
\hat{Y^{(e)}}=f(\mathcal{P}(X^{(e)},X;\theta_p);\theta),
$$
where $\mathcal{P}(X^{(e)},X;\theta_p)$ denotes the feature propagation process.
In order to make a fair comparison, we use the same linear link function $f(x;\theta)$
parameterized by $f(x;\theta)=\theta^\top x$ for all of the feature propagation processes $\mathcal{P}(.)$, and finally feed to the cross-entropy loss function:
\begin{equation}\label{eq:loss}
  \mathcal{L}=\sum_{i=1}^{m}(-y^{(e)}_{i,0}log \hat{y^{(e)}_{i,0}}-y^{(e)}_{i,1} log \hat{y^{(e)}_{i,1}})+\lambda (||W_1||^2 + ||W_2||^2).
\end{equation}
We will change the feature propagation function $\mathcal{P}(.)$ to study the performance of different types of feature propagation processes.
Specifically, we design the following two feature propagation processes in the control group, and compare with edge2vec as the treatment.


\textbf{Control1.} The first type is no feature propagation, i.e. we do not expand the edge feature at all. That is,
$$
\mathcal{P}(X^{(e)},X;\theta_p) = X^{(e)}.
$$

\textbf{Control2.} The second type is to only expand the edge feature by concatenating its source and target node features, that is,
$$\mathcal{P}(X^{(e)},X;\theta_p) = CONCAT(X^{(e)},C_sX,C_tX).$$

\textbf{Treatment (edge2vec).} The third type is to expand the edge feature by the propagation process defined in section~\ref{sec:edge}, that is,
$$
\mathcal{P}(X^{(e)},X;\theta_p)=\widetilde{X^{(e)}}
$$
where $\widetilde{X^{(e)}}$ is computed by Eq.~\eqref{eq:feaprop_reduce}.

\subsection{Results and Analyses}
\begin{figure} [!th]
\centering\vspace{-0.3cm}
\includegraphics[width=70mm]{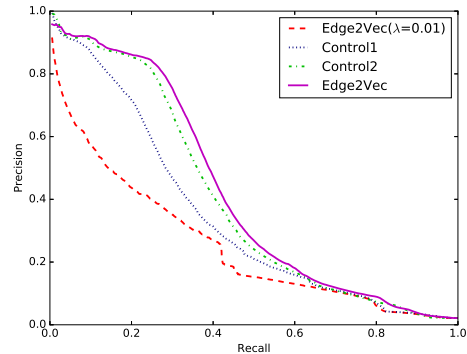}\vspace{-0.38cm}
\caption{Precision-Recall Curve}\label{fig:results}\vspace{-0.5cm}
\end{figure}

We plot the PR-curves~\footnote{\url{https://en.wikipedia.org/wiki/Precision_and_recall}} of comparison approaches in
Figure~\ref{fig:results}. We could see that the result of the treatment method edge2vec with an appropriate $\lambda$ performs much better than Control1 and a little better than Control2. Although the gain between treatment and Control2 is not such significant, it is in line with our expectations that feature propagation could improve the performance of prediction model.

We also analyze the potential numerical issues by testing the structure2vec method~\cite{dai2016discriminative}. For the structure2vec method, its loss function introduced a penalty parameter $\lambda$ to constrain the value of $W_1$ and $W_2$. If $\lambda$ is set up as a small value, the numerical issue will rise up. We take four numbers of $\lambda$ $\{10^{-3},10^{-4},10^{-5},10^{-6}\}$
and then test in which order of steps $\{1-order,2-orders,3-orders,4-orders,5-orders\}$ (see into section~\ref{sec:relationship1}) will lead to numerical overflow.

The following table displays the test results. We can find out that when $\lambda$ becomes small
enough($10^{-5}$ or $10^{-6}$), the numerical overflow issue happens. However, setting a large $\lambda$ is not a good method to handle this issue, for a large $\lambda$ may weaken the model's performance very sharply (see the curve of edge2vec under $\lambda=0.01$ in Figure~\ref{fig:results}.

\begin{table}[!th]
\centering
\vspace{-0.2cm}
\caption{Numeric Overflow (Overflow or Not).} \label{tab:data}
\vspace{-0.3cm}
\begin{small}
  \begin{tabular}{|c|c|c|c|c|c|c|}
\hline
 &1-order&2-orders&3-orders&4-orders&5-orders \\\hline
$10^{-3}$ &N& N& N& N& N \\ \hline
$10^{-4}$ &N& N& N& N& N \\ \hline
$10^{-5}$ &N& Y& Y& Y& Y \\ \hline
$10^{-6}$ &Y& Y& Y& Y& Y \\ \hline
\end{tabular}
\vspace{-0.6cm}
\end{small}
\end{table}

\section{Conclusion and Future Work}
In this paper, we proposed a new concept ``feature propagation'' and a typical way for feature propagation. We proved that convergence is a noteworthy issue for feature propagation and proposed certain conditions to guarantee its convergence. Then we revisited the two types of graph representation learning methods and found both of them have strong connections with feature propagation. Although we only revisited very limited graph representation learning methods, we provided a new perspective for understanding the essence of graph representation learning. The experiment on fraud transaction detection demonstrated the method with feature propagation could do better than the method without it. We also tested the numerical overflow issue in structure2vec. It's a pity that we only pointed out the issue but haven't proposed a practical way to make it. We think it is a worthy direction to explore in the future.

\bibliographystyle{named}

\begin{thebibliography}{}

\bibitem[\protect\citeauthoryear{Abu-El-Haija \bgroup \em et al.\egroup
  }{2017}]{abu2017watch}
Sami Abu-El-Haija, Bryan Perozzi, Rami Al-Rfou, and Alex Alemi.
\newblock Watch your step: Learning graph embeddings through attention.
\newblock {\em arXiv preprint arXiv:1710.09599}, 2017.

\bibitem[\protect\citeauthoryear{Belkin and Niyogi}{2002}]{belkin2002laplacian}
Mikhail Belkin and Partha Niyogi.
\newblock Laplacian eigenmaps and spectral techniques for embedding and
  clustering.
\newblock In {\em Advances in neural information processing systems}, pages
  585--591, 2002.

\bibitem[\protect\citeauthoryear{Cao \bgroup \em et al.\egroup
  }{2015}]{cao2015grarep}
Shaosheng Cao, Wei Lu, and Qiongkai Xu.
\newblock Grarep: Learning graph representations with global structural
  information.
\newblock In {\em Proceedings of the 24th ACM International on Conference on
  Information and Knowledge Management}, pages 891--900. ACM, 2015.

\bibitem[\protect\citeauthoryear{Dai \bgroup \em et al.\egroup
  }{2016}]{dai2016discriminative}
Hanjun Dai, Bo~Dai, and Le~Song.
\newblock Discriminative embeddings of latent variable models for structured
  data.
\newblock In {\em International Conference on Machine Learning}, pages
  2702--2711, 2016.

\bibitem[\protect\citeauthoryear{Grover and
  Leskovec}{2016}]{grover2016node2vec}
Aditya Grover and Jure Leskovec.
\newblock node2vec: Scalable feature learning for networks.
\newblock In {\em Proceedings of the 22nd ACM SIGKDD international conference
  on Knowledge discovery and data mining}, pages 855--864. ACM, 2016.

\bibitem[\protect\citeauthoryear{Hamilton \bgroup \em et al.\egroup
  }{2017}]{hamilton2017inductive}
William~L Hamilton, Rex Ying, and Jure Leskovec.
\newblock Inductive representation learning on large graphs.
\newblock {\em arXiv preprint arXiv:1706.02216}, 2017.

\bibitem[\protect\citeauthoryear{Kipf and Welling}{2016}]{kipf2016semi}
Thomas~N Kipf and Max Welling.
\newblock Semi-supervised classification with graph convolutional networks.
\newblock {\em arXiv preprint arXiv:1609.02907}, 2016.

\bibitem[\protect\citeauthoryear{Langville and
  Meyer}{2006}]{langville2006updating}
Amy~N Langville and Carl~D Meyer.
\newblock Updating markov chains with an eye on google's pagerank.
\newblock {\em SIAM journal on matrix analysis and applications},
  27(4):968--987, 2006.

\bibitem[\protect\citeauthoryear{Liu \bgroup \em et al.\egroup
  }{2017}]{Liu:2017:PNN:3133956.3138827}
Ziqi Liu, Chaochao Chen, Jun Zhou, Xiaolong Li, Feng Xu, Tao Chen, and Le~Song.
\newblock Poster: Neural network-based graph embedding for malicious accounts
  detection.
\newblock In {\em Proceedings of the 2017 ACM SIGSAC Conference on Computer and
  Communications Security}, CCS '17, pages 2543--2545, New York, NY, USA, 2017.
  ACM.

\bibitem[\protect\citeauthoryear{Page \bgroup \em et al.\egroup
  }{1999}]{page1999pagerank}
Lawrence Page, Sergey Brin, Rajeev Motwani, and Terry Winograd.
\newblock The pagerank citation ranking: Bringing order to the web.
\newblock Technical report, Stanford InfoLab, 1999.

\bibitem[\protect\citeauthoryear{Pennington \bgroup \em et al.\egroup
  }{2014}]{pennington2014glove}
Jeffrey Pennington, Richard Socher, and Christopher Manning.
\newblock Glove: Global vectors for word representation.
\newblock In {\em Proceedings of the 2014 conference on empirical methods in
  natural language processing (EMNLP)}, pages 1532--1543, 2014.

\bibitem[\protect\citeauthoryear{Perozzi \bgroup \em et al.\egroup
  }{2014}]{perozzi2014deepwalk}
Bryan Perozzi, Rami Al-Rfou, and Steven Skiena.
\newblock Deepwalk: Online learning of social representations.
\newblock In {\em Proceedings of the 20th ACM SIGKDD international conference
  on Knowledge discovery and data mining}, pages 701--710. ACM, 2014.

\bibitem[\protect\citeauthoryear{Plemmons}{1977}]{plemmons1977m}
Robert~J Plemmons.
\newblock M-matrix characterizations. i—nonsingular m-matrices.
\newblock {\em Linear Algebra and its Applications}, 18(2):175--188, 1977.

\bibitem[\protect\citeauthoryear{Sen \bgroup \em et al.\egroup
  }{2008}]{sen2008collective}
Prithviraj Sen, Galileo Namata, Mustafa Bilgic, Lise Getoor, Brian Galligher,
  and Tina Eliassi-Rad.
\newblock Collective classification in network data.
\newblock {\em AI magazine}, 29(3):93, 2008.

\bibitem[\protect\citeauthoryear{Xiang \bgroup \em et al.\egroup
  }{2013}]{xiang2013pagerank}
Biao Xiang, Qi~Liu, Enhong Chen, Hui Xiong, Yi~Zheng, and Yu~Yang.
\newblock Pagerank with priors: An influence propagation perspective.
\newblock In {\em IJCAI}, pages 2740--2746, 2013.

\bibitem[\protect\citeauthoryear{Zitnik and
  Leskovec}{2017}]{zitnik2017predicting}
Marinka Zitnik and Jure Leskovec.
\newblock Predicting multicellular function through multi-layer tissue
  networks.
\newblock {\em Bioinformatics}, 33(14):i190--i198, 2017.

\end{thebibliography}

\end{document}